\documentstyle[11pt]{article}
\begin{document}
\begin{center}
{\Large \bf Supersymmetry of Noncompact MQCD-like Membrane Instantons and Heat Kernel
Asymptotics} \vskip 0.1in
{Kanishka Belani\footnote{email: imkani@hotmail.com}, Payal Kaura\footnote{email: pa123dph@iitr.ernet.in} and Aalok Misra\footnote{e-mail: aalokfph@iitr.ernet.in}\\
Indian Institute of Technology Roorkee,\\
Roorkee - 247 667, Uttaranchal, India}
\vskip 0.5 true in
\end{center}

\begin{abstract}
We perform a heat kernel asymptotics analysis of the nonperturbative
superpotential obtained from wrapping of an $M2$-brane around a
supersymmetric noncompact three-fold embedded in a (noncompact)
$G_2$-manifold as obtained in \cite{Vol2}, the three-fold being the
one relevant to domain walls in Witten's MQCD \cite{MQCD1}, in the
limit of small ``$\zeta$", a complex constant that appears in the
Riemann surfaces relevant to defining the boundary conditions for
the domain wall in MQCD. The MQCD-like configuration is
interpretable, for small but non-zero $\zeta$ as a
noncompact/``large" open membrane instanton, and for vanishing
$\zeta$, as the type IIA D0-brane (for vanishing $M$-theory circle
radius). We find that the eta-function Seeley de-Witt coefficients
vanish, and we get a perfect match between the zeta-function Seeley
de-Witt coefficients (up to terms quadratic in $\zeta$) between the
Dirac-type operator and one of the two Laplace-type operators
figuring in the superpotential. Given the dissimilar forms of the
bosonic and the square of the fermionic operators, this is an
extremely nontrivial check, from a spectral analysis point of view,
of the expected residual supersymmetry for the nonperturbative
configurations in $M$-theory considered in this work.
\end{abstract}

\newpage

\section{Introduction}

String and $M$ theories on manifolds with $G_2$ and $Spin(7)$ holonomies
have become an active area of research, after construction of explicit
examples of such manifolds by Joyce\cite{J}. Some explicit metrics
of noncompact manifolds with the above-mentioned exceptional holonomy groups
have been constructed by Brandhuber et al and Cvectic et al
\cite{C}.

Further, it will be interesting to be able to lift the
Gopakumar-Vafa Chern-Simons/closed type A-topological open/closed
string duality to $M$ theory on a $G_2$-holonomy manifold, without
having to embed it first into type IIA string theory as was done by
Vafa. As the type-A topological string theory's partition function
receives contributions only from holomorphic maps from the
world-sheet to the target space, and apart from constant maps,
instantons fit the bill, as a first step we should look at obtaining
the superpotential contribution of wrapping of an $M2$ brane on
supersymmetric 3-cycle in a suitable $G_2$-holonomy
manifold(membrane instantons). In terms of relating the result
obtained in \cite{AM}(on membrane instanton superpotential in terms
of bosonic and fermionic determinants) to that of the 1-loop
Schwinger computation of $M$ theory and the large $N$-limit of the
partition function evaluated in \cite{GopVafaMth1}, one notes that
the 1-loop Schwinger computation also has as its starting point, an
infinite dimensional bosonic determinant of the type
$det\biggl((i\partial-e A)^2 -Z^2\biggr)$, $A$ being the gauge field
corresponding to an external self-dual field strength, and $Z$
denoting the central charge. The large $N$-limit of the partition
function of Chern Simons theory on an $S^3$, as first given by
Periwal in \cite{Periwal}, involves the product of infinite number
of $sin$'s, that can be treated as the eigenvalues of an infinite
dimensional determinant. This is indicative of a possible connection
between the membrane instanton contribution to the superpotential,
the 1-loop Schwinger computation and the large $N$ limit of the
Chern-Simons theory on an $S^3$ (See also \cite{Curio}).

After the construction of noncompact (special Lagrangian)
three-folds of Joyce \cite{NCSLAGS} given by the following equations
(noncompact SLAGS in ${\bf C}^3$):
$|z_i|^2-t=|z_j|^2=|z_k|^2, {\rm Re}(z_1z_2z_3)\geq0,\
{\rm Im}(z_1z_2z_3)=0, t>0,\ i\neq j\neq k=1,2,3$,
the same have been studied in the context of wrapping of $D6$-branes
around noncompact SLAGS diffeomorphic to $S^1\times{\bf R}^2$
in \cite{Kachruetal}
- see also \cite{Klemmetal}. In this paper, we identify a noncompact
instantonic configuration in M-theory compactified on a
$G_2$-manifold with a particular supersymmetric noncompact
three-fold embedded in the same, as relevant to domain walls in
MQCD. What this means is that we consider
$M$ theory compactified on the same $G_2$-manifold with a supersymmetric
noncompact three-fold embedded in it that appears in the study of
domain walls in MQCD, but we do {\it not} work with MQCD - we just ``borrow''
the MQCD-domain-wall $G_2$, but continue working with $M$ theory which gives
${\cal N}=1, D=4$ chiral multiplets (in addition to gravity and $U(1)$ vector
multiplets) - there are no chiral multiplets in ${\cal N}=1$ MQCD. We are
considering the superpotential of an isolated membrane instanton obtained
by wrapping of an $M2$-brane on the aforementioned supersymmetric three-fold.
We provide evidence for the expected residual supersymmetry
for the same, from a spectral analysis point of view, by looking at
the Seeley-de Witt coefficients associated with the Laplace-type and
Dirac-type operators relevant to the nonperturbative superpotential
for small value of a complex constant ``$\zeta$" that figures in
Riemann surfaces relevant to MQCD. Notice that the reason why from
the heat kernel asymptotics' point of view, the fact that the
instantonic configuration possesses some surviving supersymmetry is
not obvious is because if one looks at the forms of the Laplace-type
(bosonic) operator (the relevant one denoted in this paper by ${\cal
O}_1$) and the Dirac-type (fermionic) operator (denoted in this
paper by ${\cal O}_3$), then writing ${\cal
O}_1=G^{ij}\partial_i\partial_j + X_1$ and ${\cal
O}_3^2=G^{ij}\partial_i\partial_j + X_2$, one sees that $X_1$ is not
the same as $X_2$. Had $X_1$ equalled $X_2$, then using a theorem of
McKean and Singer (\cite{MS}), one would have been assured of
absence of UV divergence in ${\rm ln\ det} {\cal O}_1 - {\rm ln\
det} {\cal O}_3$. The fact that, as we will show in section {\bf 4},
one still gets a perfect cancelation (to the order considered in our
work), is, we feel, a very non-trivial result.

The main idea behind this paper is to use the heat kernel
asymptotics techniques to see whether or not the aforementioned
membrane instanton superpotential receives quantum corrections from
the bosonic and fermionic fluctuations, thereby verifying the
expected (because one is using a supersymmetric three-fold for
wrapping the $M2$-brane) surviving supersymmetry. Further, as far as we know, a spectral
analysis for supersymmetric three-folds embedded in a $G_2$-manifold, has never been done
earlier.

The plan of the paper is as follows. In section {\bf 2}, we review
the calculation of the membrane instanton superpotential of
\cite{AM}, which is based on \cite{Ovrutetal} - for this work, given the
non-singular uplift to M-theory in MQCD, we assume that there is no
contribution to the superpotential from the membrane
boundary. In section {\bf 3}, we discuss spin connections for
associative three-cycles based on the discussion on the same in
\cite{harvmoore}. In section {\bf 4}, we perform a heat kernel
asymptotics analysis for the superpotential of section {\bf 2} and
using the results of \cite{Gilkey}, evaluate the bulk and boundary
Seeley de-Witt coefficients for one of the two Laplace-type
operators and the Dirac-type operator. We obtain a remarkably perfect match
between the two (up to ${\cal O}(\zeta^2)$) thereby strongly indicative of
surviving supersymmetry of the nonperturbative configurations in $M$-theory considered in this work.

\section{Evaluation of the membrane instanton contribution to the superpotential}

In \cite{AM}, one of us (AM) had worked out the membrane instanton
superpotential, using techniques developed in \cite{Ovrutetal},
based on the path-integral-inside-a-path integral approach of
\cite{BBS}. We briefly review the same first.

As given in \cite{harvmoore}, the Euclidean action for an $M2$ brane is
given
by the following Bergshoeff, Sezgin, Townsend action:
\begin{equation}
\label{eq:EucM2} {\cal S}_\Sigma=\int_\Sigma
d^3z\Biggl[{\sqrt{g}\over l_{11}^3} -{i\over 3!}\epsilon^{ijk}
\partial_i{\bf Z}^M\partial_j{\bf Z}^N\partial_k{\bf Z}^PC_{MNP}
(X(s),\Theta(s))\Biggr],
\end{equation}
where ${\bf Z}$ is the map of the $M2$ brane world-volume to the the
$D=11$ target space $M_{11}$, both being regarded as supermanifolds
and $\Sigma$ is the $M2$-brane world volume. The $g$ in
(\ref{eq:EucM2}), is defined as:
\begin{equation}
\label{eq:gdef}
g_{ij}=\partial_i{\bf Z}^M\partial_j{\bf Z}^N{\bf E}^A_M{\bf
E}^B_N\eta_{AB},
\end{equation}
where ${\bf E}^A_M$ is the supervielbein, given in \cite{harvmoore}.
$X(s)$ and $\Theta(s)$ are the bosonic and fermionic coordinates
of ${\bf Z}$. After using the static gauge and
$\kappa$-symmetry fixing, the physical degrees of freedom, are given
by $y^{m^{\prime\prime}}$, the section of the normal bundle to
the $M2$-brane world volume, and $\Theta(s)$, section of the
spinor bundle tensor product: $S(T\Sigma)\otimes S^-(N)$, where
the $-$ is the negative $Spin(8)$ chirality, as under an orthogonal
decomposition of $TM_{11}|_\Sigma$ in terms of tangent and normal
bundles, the structure group $Spin(11)$ decomposes into
$Spin(3)\times Spin(8)$.

The action in (\ref{eq:EucM2}) needs to be expanded up to $O(\Theta^2)$,
and the expression is (one has to be careful that in Euclidean $D=11$, one does
not have a Majorana-Weyl spinor or a Majorana spinor) given as:
\begin{eqnarray}
\label{eq:actexpTh2} & & {\cal S}_\Sigma=\int_\Sigma\Biggl[C
+{i\over l_{11}^3}vol(g) +{\sqrt{g}\over l_{11}^3}\biggl(g^{ij}
D_iy^{m^{\prime\prime}} D_j y^{n^{\prime\prime}}
h_{m^{\prime\prime}n^{\prime\prime}}-y^{m^{\prime\prime}} {\cal
U}_{m^{\prime\prime}n^{\prime\prime}}y^{n^{\prime\prime}}+O(y^3)
\biggr)\nonumber\\
& & +{i\over l_{11}^3}\sqrt{g}{1\over2}({\bar\Psi}_MV^M
- {\bar V}^M\Psi_M)+2{\sqrt{g}\over l_{11}^3}g^{ij}
{\bar\Theta}\Gamma_iD_j\Theta + O(\Theta^3)\Biggr],
\end{eqnarray}
where we follow the conventions of \cite{harvmoore}: $V_M$ being the
gravitino vertex operator, $\Psi$ being the gravitino field that
enters via the supervielbein ${\bf E}^A_M$, ${\cal U}$ is a mass
matrix defined in terms of the Riemann curvature tensor and the
second fundamental form, $C$ is the pull-back of the $M$-theory
three-form potential on to the world volume of the $M2$-brane and
$\Gamma_i$s are pull-backs of
the eleven-dimensional gamma matrices on to $\Sigma$.

After $\kappa$-symmetry fixing, like \cite{harvmoore}, we set
$\Theta_2^{A\stackrel{.}{a}}(s)$ ($A$ and $\stackrel{.}{a}$ index
the $Spin(3)$ and the positive-chirality $Spin(8)$ groups
respectively), i.e., the positive $Spin(8)$-chirality, to zero, and
following \cite{Ovrutetal}, will refer to $\Theta_1^{Aa}(s)$ as
$\theta$.

The Kaluza-Klein reduction of the $D=11$ gravitino $\Psi_M$, is
given by: $dx^M\Psi_M=dx^\mu\Psi_\mu+dx^m\Psi_\mu,$ ($\mu$ indexes
the four-dimensional Euclidean space ${\bf R}^4(x)$ and $m$ indexes
the $G_2$ seven-fold
$X_{G_2}(y)$)$\Psi_\mu(x,y)=\psi_\mu(x)\otimes\vartheta(y)$,
$\Psi_m(x,y)=l_{11}^3\sum_{i=1}^{b_3}
\omega^{(3)}_{i,mnp}(y)\Gamma^{pq}\chi^i(x)\otimes\eta(y)$,
$\omega^{(3)}\in H^3(X_{G_2})$, where we do not write the terms
obtained by expanding in terms of a basis of the harmonic 2-forms of
$H^2(X_{G_2})$, as we will be interested in $M2$ branes wrapping
supersymmetric 3-cycles in the $G_2$-holonomy manifold -
$\Gamma^{pq}$ is the antisymmetrized product of two $Cl(0,11)$
generators, $\psi_\mu$ is the four-dimensional gravitino in the
four-dimensional ${\cal N}=1$ (super)gravity multiplet, $\chi^i$ are
the four-dimensional fermions in the ${\cal N}=1$ chiral multiplet
($M$-theory compacitified on a $G_2$ manifold would yield a
four-dimensional ${\cal N}=1$ theory) and $\eta(y)$ is a covariantly
constant spinor on the $G_2$-manifold. For evaluating the
nonperturbative contribution to the superpotential, following
\cite{harvmoore}, we will evaluate the fermionic 2-point function:
$\langle\chi^i(x_1^u)\chi^j(x_2^u)\rangle$ (where $x_{1,2}$ are the
${\bf R}^4$ coordinates and u [and later also $v$]$\equiv 7,8,9,10$
is [are] used to index these coordinates), and drop the interaction
terms in the $D=4,{\cal N}=1$ supergravity action. The corresponding
mass term in the supergravity action appears as
$\partial_i\partial_j W$, where the derivatives are evaluated w.r.t.
the complex scalars obtained by the Kaluza-Klein reduction of
$C+{i\over l_{11}^3}\Phi$ ($\Phi$ being the closed as well as
co-closed $G_2$-calibration three form defined over $X_{G_2}$) using
harmonic three forms forming a basis for $H^3(X_{G_2},{\bf R})$. One
then integrates twice to get the expression for the superpotential
from the 2-point function.

The bosonic zero modes are the four bosonic coordinates that specify
the position of the supersymmetric 3-cycle, and will be denoted
by $x_0^{7,8,9,10}\equiv x^u_0$. The fermionic zero modes come from
the fact that for every $\theta_0$ that is the solution to the fermionic
equation of motion, one can always shift $\theta_0$ to
$\theta_0+\theta^\prime$
, where $D_i\theta^\prime=0$. This $\theta^\prime=\vartheta\otimes\eta$
where $\vartheta$ is a $D=4$ Weyl spinor, and $\eta$ is a covariantly
constant spinor on the $G_2$-holonomy manifold.

After expanding the $M2$-brane action in fluctuations about solutions to
the bosonic and fermionic equations of motion, one gets that:
${\cal S}|_\Sigma={\cal S}^y_0+{\cal S}^\theta_0+{\cal S}^y_2
+{\cal S}^\theta_2$,
where
${\cal S}^y_0\equiv {\cal S}_\Sigma|_{y_0,\theta_0}$
${\cal S}^\theta_0\equiv {\cal S}_\Sigma^\theta
+{\cal S}^{\theta^2}_\Sigma|_{y_0,\theta_0}$;
${\cal S}^y_2
\equiv {\delta^2{\cal S}_\Sigma\over\delta y^2}|_{y_0,\theta_0=0}
(\delta y)^2$;
${\cal S}^\theta_2
\equiv {\delta^2{\cal S}\over\delta\theta^2}|_{y_0,\theta_0=0}
(\delta\theta)^2$.
Following \cite{Ovrutetal}, we consider
classical values of coefficients of $(\delta y)^2,(\delta\theta)^2$ terms,
as fluctuations are considered to be of ${\cal O}(\sqrt{\alpha^\prime})$.

Now,
\begin{eqnarray}
\label{eq:2ptdef}
& & \langle \chi^i(x_1^u)\chi^j(x_2^u)\rangle=\nonumber\\
& & \int {\cal D}\chi e^{K.E\ of\ \chi}\chi^i(x)\chi^j(x)
\int d^4x_0 e^{-{\cal S}_0^y}\nonumber\\
& & \times\int d\vartheta^1d\vartheta^2 e^{-{\cal S}^\theta_0}
\int{\cal D}\delta y^{m^{\prime\prime}}e^{-{\cal S}^y_2}
\int{\cal D}\delta{\bar\theta}{\cal D}\delta\theta
e^{-{\cal S}^\theta_2}.
\end{eqnarray}

We now evaluate the various integrals that appear in (\ref{eq:2ptdef}) above
starting with $\int d^4x e^{-{\cal S}_0^y}$:
\begin{equation}
\label{eq:bzmint}
\int d^4x_0 e^{-{\cal S}_0^y}=\int d^4x_0 e^{[iC-{1\over l_{11}^3}vol(g)]}.
\end{equation}

Using the 11-dimensional Euclidean representation of the gamma matrices
as given in \cite{harvmoore},
${\cal S}^\theta_0+{\cal S}^{\theta^2}_0|_\Sigma={i\over
2l_{11}^3}\int_\Sigma
\sqrt{g}{\bar\Psi}_MV_M d^3s,$
where using $\partial_i x_0^u=0$, and using $U$ to denote coordinates
on the $G_2$-holonomy manifold, $V^U=g_{ij}\partial_i y_0^U\partial_j y^V
\gamma_V\theta_0+{i\over2}\epsilon^{ijk}\partial_iy_0^U
\partial_jy_0^V\partial_k y_0^W\Gamma_{VW}\theta_0$,
\begin{equation}
\label{eq:linth}
\int d\vartheta_1 d\vartheta_2e^{{i\over 2l_{11}^3}\sum_{I=1}^{b_3}
\sum_{\alpha=1}^2\sum_{i=1}^8
({\bar\chi}(x)\sigma^{(i)})_\alpha\vartheta_\alpha\omega_I^{(i)}}
=-{1\over 4l_{11}^3}\sum_{I=1}^{b_3}\sum_{i<j=1}^8\omega_I^{(i)}
\omega_I^{(j)}({\bar\chi}\sigma^{(i)})_1({\bar\chi}\sigma^{(j)})_2,
\end{equation}
where one uses that for $G_2$-spinors, the only non-zero bilinears
are: $\eta^\dagger\Gamma_{i_1...i_p}\eta$ for $p=0(\equiv$
constant), $p=3(\equiv$ calibration 3-form), $p=4(\equiv$ Hodge dual
of the calibration 3-form) and $p=7(\equiv$ volume form). We follow
the following notations for coordinates: $u,v$ are ${\bf R}^4$
coordinates and $U,V$ index $G_2$-holonomy manifold coordinates. The
tangent/curved space coordinates for $\Sigma$ are represented by
$a^\prime/m^\prime$ and those for $X_{G_2}\times{\bf R}^4$ are
represented by $a^{\prime\prime}/m^{\prime\prime}$.

We now come to the evaluation of ${\cal
S}^\theta_2|_{y_0,\theta_0=0}$. Using the equality of the two
$O((\delta\Theta)^2)$ terms in the action of Harvey and Moore, and
arguments similar to the ones in \cite{Ovrutetal}, one can show that
one needs to evaluate the following bilinears:
$\delta{\bar\Theta}\Gamma_{a^\prime}\partial_i\delta\Theta$,
$\delta{\bar\Theta}\Gamma_{a^{\prime\prime}}\partial_i\delta\Theta$,
$\delta{\bar\Theta}\Gamma_{a^\prime}\Gamma_{AB}\delta\Theta$, and
$\delta{\bar\Theta}\Gamma_{m^{\prime\prime}}\Gamma_{AB}\delta\Theta$.
Evaluating them, one gets:
\begin{equation}
\label{eq:th2} {\cal S}^\theta_2|_{y_0,\theta_0=0}\equiv\int_\Sigma
d^3s \delta \theta^\dagger{\cal O}_3\delta\theta,
\end{equation}
where ${\cal O}_3\equiv\sqrt{g}g^{ij}\Gamma_iD_j$, the precise definition of $\Gamma_i$ will be given later.
Hence, the integral over the fluctuations in $\theta$ will give a
factor of $\sqrt{det{\cal O}_3}$ in Euclidean space.

The expression for ${{\cal S}^y_2|_\Sigma}_{y_0,\theta_0=0}$ is identical
to the one given in \cite{Ovrutetal}, and will contribute
${1\over{\sqrt{det{\cal O}_1det{\cal O}_2}}}$, where ${\cal O}_1$ and
${\cal O}_2$ are as given in the same paper:
\begin{eqnarray}
\label{eq:bosondets}
& & {\cal O}_1\equiv\eta_{uv}\sqrt{g}g^{ij}{\cal D}_i\partial_j\nonumber\\
& & {\cal O}_2\equiv\sqrt{g}(g^{ij}{\cal D}_ih_{\hat{U}\hat{V}}
D_j+{\cal U}_{\hat{U}\hat{V}}).
\end{eqnarray}
The mass matrix ${\cal U}$ is expressed in terms of the curvature
tensor and product of two second fundamental forms. ${\cal D}_i$ is
a covariant derivative with indices in the corresponding
spin-connection of the type
$(\omega_i)^{m^{\prime\prime}}_{n^{\prime\prime}}$ and
$(\omega_i)^{m^\prime}_{n^\prime}$, and $D_i$ is a covariant
derivative with corresponding spin connection indices of only the
latter type.

Hence, modulo supergravity determinants, and the contribution from the
fermionic zero modes, the exact form of the superpotential
contribution coming from a single $M2$ brane wrapping an isolated supersymmetric
cycle of $G_2$-holonomy manifold, is given by:
\begin{equation}
\label{eq:Wfinal} \Delta W = e^{iC - {1\over
l_{11}^3}vol(h)}\sqrt{{det{\cal O}_3\over det {\cal O}_1\ det{\cal
O}_2}}.
\end{equation}
Note that the result (\ref{eq:Wfinal}), unlike that of
\cite{harvmoore}, is also applicable for non-rigid three-cycles
(implying $b_1(\Sigma)\neq0$). We do not bother about 5-brane
instantons, as we assume that $H_6=0$ for the $G_2$-manifold. One
should bear in mind that it is only for {\it compact}
$G_2$-manifolds $X_7$ that $H^4_7(X_7)$, valued in the
seven-dimensional representation of the $G_2$-group, and therefore
$H_3(X_7)$ vanishes - note however $H^4(X_7)=H^4_1(X_7)\oplus
H^4_7(X_7)\oplus H^4_{27}(X_7)$ (for a compact $X_7$) - hence (for a
compact $X_7$), $H^4(X_7)$ and hence $H_3(X_7)$ is non-trivial.
Besides, we are working with a noncompact $G_2$ manifold
\cite{Joyce1}. Even though we have turned off the $G$-flux and the
calibration three-form characterizing $X_7$ is closed, $H^4(X_7)$
and therefore $H_3(X_7)$ are still non-trivial. Part of the reason
is the shift of the quantization of the $G$-flux (See
\cite{Witten}): $\left[\frac{G}{2\pi}\right]-\frac{\lambda}{2}\in
H^4(X_7,{\bf Z})$, where the characteristic class $\lambda$ is given
by $\frac{p_1(X_7)}{2}$, $p_1(X_7)$ being the first Pontryagin
class. For a $G_2$ manifold, $p_1(X_7)=p_1(X_8=X_7\times S^1)$,
$X_7\times S^1$ being a spin eight-manifold (See \cite{harvmoore}),
for which $p_1$ is even. According to the Wu's formula (See
\cite{Witten}), the intersection form of a spin eight-fold satisfies
the following relation: $x^2\cong x\wedge\lambda\ {\rm mod}\ 2$,
where $x\in H^4(X_8,{\bf Z})$. Hence, for a spin manifold, $\lambda$
is even if the intersection form is even - the intersection form for
$X_8$ is even, thereby justifying the switching off of $G$. Further,
$p_1(X_7)\neq0$\footnote{As an example of a compact $X_7$, one could
consider $\frac{T^7}{\Gamma}$ with fixed points for $\Gamma$ - see
\cite{J} - $p_1(X_7)$ for compact $X_7$ with holonomy given by $G_2$,
is non-zero and satisfies the equation: $\langle p_1\cup\phi,
[X_7]\rangle=-\frac{1}{8\pi^2}\int_{X_7}|R|^2<0$ (See \cite{Joyce1}), $\phi$
being the (co)closed calibration three-form}
implying that $H^4(X_7)$ and therefore $H_3(X_7)$ can not be
trivial.

To actually evaluate the Seeley-de Witt coefficients for
Laplace-type operators (See (\ref{eq:bosSdW})) we need to find an
example of a regular $G_2$-holonomy manifold that is locally
$\Sigma\times M_4$, where $\Sigma$ is a supersymmetric 3-cycle on
which we wrap an $M2$ brane once, and $M_4$ is a four manifold. The
condition for supersymmetric cycle: $\Phi|_\Sigma={\rm
vol}(\Sigma)$, is what is solved for in \cite{Vol2}. Such a
three-fold will be discussed in section {\bf 4} in the context of
MQCD.

\section{Spin Connection for Associative 3-Cycles in $G_2$ Manifolds}

We now discuss how to figure out the independent components of the
spin connection of the type $\omega_i^{a^\prime
b^{\prime\prime}}$(the superindices indexing the tangent space
indices as explained below equation (6) above: $i$ indexes $\Sigma$,
$a^\prime$ indexes the tangent space of $\Sigma$ and
$b^{\prime\prime}$ indexes the tangent space corresponding to
directions normal to $\Sigma$).

The bi-spinorial representation of the components of the spin
connection $\omega^{a^\prime b^\prime}_i, \omega^{a^{\prime\prime}
b^{\prime\prime}}_i, \omega^{a^\prime b^{\prime\prime}}_i$ can be
worked out as below. Let $\omega^{a^\prime b^\prime}_i\equiv
\epsilon^{a^\prime b^\prime c^\prime}\omega^{c^\prime}_i$. We can
then construct
$\omega^{AB}_i\equiv(\sum_{a^\prime=1}^3\omega^{a^\prime+2}_i
\sigma^{a^\prime})^{AB}\equiv\omega^{AB}_i$ (where
$\sigma^{a^\prime}$ are the Pauli matrices and $A,B$ are the
bispinorial indices), abbreviated as $\omega_\parallel$. The
components $\omega^{a^{\prime\prime} b^{\prime\prime}}_i$, with
$a^{\prime\prime},b^{\prime\prime}=6,7,8,10$ can be split into three
self-dual and three anti-self-dual components:
\begin{eqnarray}
\label{eq:6to33splitSD}
& & \omega^{67}_i+\omega^{8 10}_i\equiv(\omega_i^+)^1,\nonumber\\
& & \omega^{68}_i+\omega^{10\ 7}_i\equiv(\omega_i^+)^2,\nonumber\\
& & \omega^{6\ 10}_i+\omega^{78}_i\equiv(\omega_i^+)^3,
\end{eqnarray}
and
\begin{eqnarray}
\label{eq:6to33splitASD}
& & \omega^{67}_i-\omega^{8 10}_i\equiv(\omega_i^-)^1,\nonumber\\
& & \omega^{68}_i-\omega^{10\ 7}_i\equiv(\omega_i^-)^2,\nonumber\\
& & \omega^{6\ 10}_i-\omega^{78}_i\equiv(\omega_i^-)^3,
\end{eqnarray}
from which one constructs $\sum_{a=1}^3(\omega^+_i)^a\sigma^a\equiv
(\omega^+_i)^{YY^\prime}$, abbreviated as $\omega_\perp^+$, and
$\sum_{a=1}^3(\omega^-_i)^a\sigma^a
\equiv(\omega^-_i)^{\dot{Y}\dot{Y}^\prime}$, abbreviated as
$\omega_\perp^-$. For the ``off-diagonal" components
$\omega^{a^\prime b^{\prime\prime}}$, one constructs
$\sum_{a^\prime=1}^3(\omega^{a^\prime
b^{\prime\prime}})\sigma^{a^\prime}
\equiv(\omega_i)^{ABb^{\prime\prime}}$, and further $(
\omega^{AB6}_i{\bf 1}_2+ \omega^{AB7}_i\sigma^1+
\omega^{AB8}_i\sigma^2+ \omega^{AB\
10}_i\sigma^3)^{Y\dot{Y}}\equiv(\omega)_i^{ABY\dot{Y}}$.

For associative three-cycles, $(\omega)_i^{ABY\dot{Y}}$ is symmetric
w.r.t. $A,B$ and $Y$ and $\omega_\parallel=\omega_\perp^-$ (See
\cite{harvmoore}.). Assuming $\omega_\perp^+=0$, implying
$\omega^{67}_i=-\omega_i^{8\ 10}, \omega^{68}_i=-\omega^{10\ 7}_i,
\omega^{6\ 10}_i=-\omega^{78}_i$, the relation
$\omega_\parallel=\omega^-_\perp$ implies:
${1\over2}\omega^1_i=\omega^{67}_i,\
{1\over2}\omega^2_i=\omega^{68}_i,\ {1\over2}\omega^3_i=\omega^{6\
10}_i$. Hence,
\begin{equation}
\label{eq:spinconerp}
\omega^{a^{\prime\prime}b^{\prime\prime}}_i=\left(\begin{array}{cccc}
0 & {1\over2}\omega^1_i & {1\over2}\omega^2_i & {1\over2}\omega^3\\
-{1\over2}\omega^1_i & 0 & -{1\over2}\omega^3_i & {1\over2}\omega^2_i \\
-{1\over2}\omega^1_i & {1\over2}\omega^3_i & 0 & -{1\over2}\omega^{61}_i \\
-{1\over2}\omega^3_i & -{1\over2}\omega^2_i & -{1\over2}\omega^{61}_i & 0 \\
\end{array}\right).
\end{equation}

Now,
\begin{equation}
(\sum_{a^\prime=1}^3\omega_i^{a^\prime+2 \
b^{\prime\prime}}\sigma^{a^\prime})^{AB} =\left(\begin{array}{cc}
\omega_i^{5b^{\prime\prime}} &
\omega^{3b^{\prime\prime}}_i-i\omega^{4b^{\prime
\prime}}_i \\
\omega^{3b^{\prime\prime}}_i+i\omega^{4b^{\prime\prime}}_i &
-\omega^{5b^{\prime\prime}}_i \\
\end{array}\right)^{AB}.
\end{equation}
Hence, for $A=B=1$ or $2$, consider
\begin{eqnarray}
& & \pm\biggl(\omega^{56}_i{\bf
1}_2+\omega^{57}_i\sigma^1+\omega^{58}_i\sigma^2+
\omega^{5\ 10}_i\sigma^3\biggr)^{Y\dot{Y}}\nonumber\\
& & =\pm\left(\begin{array}{cc}
\omega^{56}_i+\omega^{5\ 10}_i & \omega^{57}_i-i\omega^{58}_i\\
\omega^{57}_i+i\omega^{58}_i & \omega^{56}_i-\omega^{5\ 10}_i\\
\end{array}\right)^{Y\dot{Y}},
\end{eqnarray}
and for $A=1, B=2$, consider:
\begin{eqnarray}
& & \biggl( (\omega^{36}_i-i\omega^{46}_i){\bf
1}_2+(\omega^{37}_i-i\omega^{47}_i)\sigma^1
+(\omega^{38}_i-i\omega^{48}_i)\sigma^2+(\omega^{3\
10}_i-i\omega^{4\ 10}_i)
\sigma^3\biggr)^{Y\dot{Y}}\nonumber\\
& & = \left(\begin{array}{cc}
(\omega^{36}_i-i\omega^{46}_i)+(\omega^{3\ 10}_i-i\omega^{4\ 10}_i)
&
(\omega^{37}_i-i\omega^{47}_i)-i(\omega^{38}_i-i\omega^{48}_i) \\
(\omega^{37}_i-i\omega^{47}_i)+i(\omega^{38}_i-i\omega^{48}_i) &
(\omega^{36}_i-i\omega^{46}_i)-(\omega^{3\ 10}_i-i\omega^{4\ 10}_i) \\
\end{array}\right)^{Y\dot{Y}}.
\end{eqnarray}
Now, $(\omega_i)^{111\dot{1}}, (\omega_i)^{111\dot{2}},
(\omega_i)^{222\dot{1}}, (\omega_i)^{222\dot{2}}$ are already
symmetric in $A,B,Y$. Now,
\begin{eqnarray}
& & (\omega_i)^{112\dot{1}}=\omega^{57}_i+i\omega^{58}_i,\nonumber\\
& & (\omega_i)^{121\dot{1}}=(\omega^{36}_i+\omega^{3\
10}_i)-i(\omega^{46}_i
+\omega^{4\ 10}_i),\nonumber\\
& & (\omega_i)^{211\dot{1}}=(\omega^{36}_i+i\omega^{3\
10}_i)+i(\omega^{46}_i +\omega^{4\ 10}_i).
\end{eqnarray}
Hence,
$\omega^{112\dot{1}}_i=\omega^{121\dot{1}}_i=\omega^{211\dot{1}}_i$
implies
\begin{equation}
\omega^{57}_i=\omega^{36}_i+\omega^{3\ 10}_i;\ \omega^{58}_i=0;\
\omega^{46}_i=-\omega^{4\ 10}_i.
\end{equation}
Similarly,
\begin{eqnarray}
& & (\omega_i)^{112\dot{2}}=\omega_i^{56}-\omega^{5\ 10}_i,\nonumber\\
& &
(\omega_i)^{121\dot{2}}=(\omega^{37}_i-\omega^{48}_i)-i(\omega^{47}_i+
\omega^{38}_i),\nonumber\\
& &
(\omega_i)^{211\dot{2}}=(\omega^{37}_i+\omega^{48}_i)-i(-\omega^{47}_i+
\omega^{38}_i),
\end{eqnarray}
and their equality would imply:
\begin{equation}
\omega^{47}_i=\omega^{38}_i=\omega^{48}_i=0;\
\omega^{37}_i=\omega^{56}_i-\omega^{5\ 10}_i,
\end{equation}
and the equality of:
\begin{eqnarray}
& &
(\omega_i)^{122\dot{1}}=(\omega^{37}_i+\omega^{48}_i)+i(\omega^{38}_i
-\omega^{47}_i),\nonumber\\
& &
(\omega_i)^{212\dot{1}}=(\omega^{37}_i-\omega^{48}_i)+i(\omega^{47}_i+
\omega^{38}_i),\nonumber\\
& & \omega^{221\dot{1}}_i=\omega^{5\ 10}_i-\omega^{56}_i,
\end{eqnarray}
implies
\begin{equation}
\omega^{37}_i=\omega^{56}_i=\omega^{5\
10}_i=\omega^{38}_i=\omega^{47}_i=0,
\end{equation}
and finally the equality of:
\begin{eqnarray}
& & (\omega_i)^{122\dot{2}}=(\omega^{36}_i-\omega^{3\
10}_i)-i(\omega^{46}_i
-\omega^{4\ 10}_i),\nonumber\\
& & (\omega_i)^{212\dot{2}}=(\omega^{36}_i
-\omega^{3\ 10}_i)+i(\omega^{46}_i-\omega^{4\ 10}_i),\nonumber\\
& & (\omega_i)^{221\dot{2}}=-\omega^{57}_i+i\omega^{58}_i,
\end{eqnarray}
implies:
\begin{equation}
\omega_i^{46}=\omega_i^{4\ 10}=\omega^{58}_i=\omega^{36}_i=0,\
\omega^{57}_i=\omega^{3\ 10}_i-\omega^{36}_i.
\end{equation}
One thus gets:
\begin{equation}
\omega^{a^\prime b^{\prime\prime}}_i=\left(\begin{array}{cccc}
\omega^{36}_i & \omega^{37}_i & \omega^{38}_i & \omega^{3\ 10}_i \\
\omega^{46}_i & \omega^{47}_i & \omega^{48}_i & \omega^{4\ 10}_i \\
\omega^{56}_i & \omega^{57}_i & \omega^{58}_i & \omega^{5\ 10}_i \\
\end{array}\right)=
\omega^{3\ 10}_i\left(\begin{array}{cccc}
0 & 0 & 0 & 1 \\
0 & 0 & 0 & 0 \\
1 & 1 & 0 & 1 \\
\end{array}\right).
\end{equation}
Hence, for associative three-folds, the number of independent
components in the off-diagonal spin-connection is one: $\omega^{3\
10}_i$. We will set it to zero. The off-diagonal spin-connection component
being set to zero towards the end of section 3, is a consequence of the fact
that for associative three-folds embedded in a $G_2$-manifold
(See \cite{harvmoore}),
as considered in our work, (i) the self-dual
piece of the connection on the normal bundle, ``$\omega^+_\perp$'', is
unconstrained, and (ii) the connection on the tangent bundle,
``$\omega_\parallel$'' gets identified with the anti-self dual connection on
normal bundle ``$\omega^-_\perp$'' - this is what was used in section {\bf 3}.
These follow from the covariant constancy of a $G_2$ spinor and describe
the decomposition of the adjoint representation of $G_2\subset{\rm Spin}(7)$
under $SO(3)\oplus SO(4)$. The aforementioned identification of spin
connections is standard to ``topological twisting'' (See \cite{harvmoore}
and references therein) - one must keep in mind that one of the results of
\cite{harvmoore} is that the low energy fluctuations of an $M2$-brane
wrapped around a supersymmetric three-fold is described by a three dimensional
topological field thory.
The results of this section will be
utilized when evaluating the Seeley de-Witt coefficients for the
Dirac-type operator ${\cal O}_3$ in the next section.

\section{Heat kernel Asymptotics of MQCD-like Supersymmetric Three-Fold embedded in $G_2$ Seven-Fold}

This section forms the core of the new results in this paper. We
begin with a discussion of supersymmetric three-folds relevant to
domain walls in MQCD and show that a certain infinite series, in the
limit of a small complex constant $\zeta$, can in fact be summed to
give a closed expression. We then proceed with the heat kernel
asymptotics analysis by discussing the evaluation of the generalized
zeta and eta function Seeley de-Witt coefficients relevant to the
Laplace-type operator ${\cal O}_1$ and the Dirac-type operator
${\cal O}_3$. The main result that we get is (a) vanishing of the
generalized eta function coefficients, and (b) a perfect match
between the generalized zeta function coefficients for ${\cal O}_1$
and ${\cal O}_3^2$ - all calculations are done up to ${\cal
O}(\zeta^2)$. The importance of these results was commented upon in
the introduction and we make more comments in the section on
conclusion. As an aside, we discuss a possible connection between
the antiholomorphic involution relevant to Joyce's construction of
``barely $G_2$ manifolds" from Calabi-Yau three-folds, and a similar
involution symmetry of the supersymmetric three-fold - this
three-fold turns out not to be a three-cyle, but a three-fold with
boundary (this necessitates the discussion of, both, bulk and
boundary Seeley de-Witt coefficients).

In MQCD \cite{MQCD1},
discrete chiral symmetry breaking results in the formation of domain
wall separating different vacua,
whose world-volume is topologically given by $R^{3}(x^{0,1,2})\times S(x^{3,4,5})$,
where $S$ is a supersymmetric three-fold embedded in a
$G_2$-manifold that is topologically ${\bf R}(x^3)\times{\bf
R}^5(x^{4,5,6,7,8})\times S^1(x^{10})$ Complexifying the
coordinates, $v=x^4+ix^5, w=x^7+ix^8, s=x^6+ix^{10}, t=e^{-s}$, the
boundary condition on $S$ is that as $x^3\rightarrow-\infty$,
$S\rightarrow{\bf R}\times\Sigma$ and as $x^3\rightarrow\infty$,
$S\rightarrow{\bf R}\times\Sigma^\prime$, where $\Sigma:w=\zeta
v^{-1},\ t=v^n$ and $\Sigma^\prime:w=e^{{2\pi i\over n}}\zeta
v^{-1}, t=v^n$. The calibration for $G_2$ manifolds can be written
as: $\Phi=e^{123}+e^{136}+e^{145}+e^{235}-e^{246}+e^{347} +e^{567}$
( $e^{ijk}\equiv e^i\wedge e^j\wedge e^k$), and then the
supersymmetric three-fold embedded in the $G_2$-manifold will be
given as: $w=w(x^3,v,{\bar v}), s=s(x^3,v,{\bar v})$. Then, defining
the embedding as $x^6=A(x,y,z), x^7=C(x,y,z), x^8=D(x,y,z),
x^{10}=B(x,y,z)$, $x,y,z$ being the $M2$-brane world-volume
coordinates, the condition for supersymmetric cycle:
$\Phi|_S=\sqrt{g}dx^3\wedge dx^4\wedge dx^5$, after further
relabeling $x^{3,4,5}$ as $z,x,y$ and after assuming:
$\partial_xA=\partial_yB,\
\partial_yA=-\partial_xB(\equiv$ Cauchy-Riemann condition),
translates to give:
\begin{eqnarray}
\label{eq:susyG2a} & &
[\partial_xA\partial_zA-\partial_y\partial_zB+\partial_xC\partial_zC
-\partial_yC\partial_zD]^2+[\partial_yA\partial_zA+\partial_xA\partial_zB
+\partial_yC\partial_zC+\partial_xC\partial_zD]^2\nonumber\\
& & =
[1+(\partial_xA)^2+(\partial_yA)^2+(\partial_xC)^2+(\partial_yC)^2]
[(\partial_zA)^2+(\partial_zB)^2+(\partial_zC)^2+(\partial_zD)^2].
\end{eqnarray}
The ansatz to solve (\ref{eq:susyG2a}) for the embedding of the
supersymmetric 3-cycle in the $G_2$-manifold, taken in \cite{Vol2}
was :
\begin{eqnarray}
\label{eq:SU2ans}
& & v=
\biggl[e^{{y_1\over2}}+\sum_{m=1}^\infty\biggl({1\over2cosh y_3}\biggr)^{2m}
f_{2m}(y_1)\biggr]e^{iy_2},\nonumber\\
& & w=-\zeta tanh y_3\biggl[e^{-{y_1\over2}}+\sum_{m=1}^\infty
\biggl({1\over2cosh y_3}\biggr)^{2m}
g_{2m}(y_1)\biggr]e^{-iy_1},\nonumber\\
& & s=-y_1-\sum_{m=1}^\infty\biggl({1\over2cosh y_3}\biggr)^{2m}
h_{2m}(y_1)-2iy_2,
\end{eqnarray}
where for the $SU(2)$ group, $f_{2m},g_{2m},h_{2m}$ can be complex,
but were taken to be real in \cite{Vol2}. The condition for getting
a supersymmetric 3-cycle implemented by ensuring that the pull-back
of the calibration $\Phi$ to the world volume of the 3-cycle is
identical to the volume form on the 3-cycle, gives recursion
relations between the coefficients $f_{2m}$ and $g_{2m}$, by setting
$h_{2m}=0$, e.g. for $m=1$, as shown in \cite{Vol2},
\begin{eqnarray}
\label{eq:recursionm=0} & & -\zeta
e^{-{y_1\over2}}\partial_1f_2+(2e^{y_1\over2} +{\zeta\over2}
e^{-{y_1\over2}})f_2-\zeta e^{{y_1\over2}}\partial_1g_2-(2\zeta^2
e^{-y_1\over2}+{\zeta\over2}e^{y_1\over2})g_2=-4\zeta^2
e^{{-y_1\over2}},
\nonumber\\
& & -(\zeta^2e^{-y_1}+4)f_2+2\zeta\partial_1g_2-(\zeta^2-\zeta)g_2=-2\zeta^2
e^{-y_1\over2}.
\end{eqnarray}
One can substitute for $f_2$ from the second equation and get a
second order differential equation for $g_2$. However, it is shown
that in the limit $\zeta\rightarrow0$, one can consistently set
$f_{2m}=h_{2m}=0,m\geq1$. Further, surprisingly, as perhaps missed
to be noticed in \cite{Vol2}, one also gets the following
differential equation for all $g_{2m}$'s, $m\geq1$:
\begin{equation}
\label{eq:zeta0}
2\partial_1g_{2m}+g_{2m}={\cal O}(\zeta)\rightarrow0,
\end{equation}
implying
\begin{equation}
\label{eq:zeta0sol} g_{2m}=e^{-x\over2},\ m\geq1.
\end{equation}
Hence, (\ref{eq:SU2ans}) becomes:
\begin{eqnarray}
\label{eq:SU2ans1}
& & v(x,z)=e^{{x\over2}+iy},\nonumber\\
& & w(x,z)=-\zeta {tanh(z) e^{-x\over2}\over1-({ sech(z)\over2})^2}
e^{-iy},
\nonumber\\
& & s(x,y)=-x-2iy.
\end{eqnarray}
One thus gets a convergent solution, unlike the case for finite
$\zeta$ as pointed out in \cite{Sonnenetal}.

We now consider an M-theory instanton obtained by wrapping a
Euclidean $M2$-brane around the supersymmetric noncompact three-fold
embedded in a $G_2$-manifold relevant to MQCD, and perform a heat
kernel asymptotics analysis for the membrane instanton
superpotential as obtained in (\ref{eq:Wfinal}) and explore the
possibility of cancelations between the bosonic and fermionic
determinants. For bosonic determinants $det A_b$, the function that
is relevant is the generalized zeta function, $\zeta(s|A_b)$, and that for fermionic determinants
$det A_f$, the function that is additionally relevant is the generalized eta function,
$\eta(s|A_f)$. The integral representation of the former involves
$Tr(e^{-tA_b})$, while that for the latter involves $Tr(Ae^{-tA^2})$
(See \cite{elizalde}):
\begin{eqnarray}
\label{eq:zetaetaMellin} & &
\zeta(s|A_b)={1\over\Gamma(2s)}\int_0^\infty dt t^{s-1}
Tr(e^{-tA_b}); \eta(s|A_f)={1\over\Gamma({s+1\over2})}\int_0^\infty
dt t^{{s+1\over2}}Tr(A_fe^{-tA_F^2}) ,
\end{eqnarray}
where to get the UV-divergent contributions, one looks at the
$t\rightarrow0$ limit of the two terms. To be more precise (See
\cite{Deseretal})
\begin{eqnarray}
\label{eq:bosfermdets}
& & ln det A_b = -{d\over ds}\zeta(s|A_b)|_{s=0}\nonumber\\
& & =-{d\over ds}\biggl({1\over\Gamma(s)}\int_0^\infty dt t^{s-1}
Tr(e^{-tA_b})\biggr)|_{s=0};\nonumber\\
& & ln det A_f = -{1\over2}{d\over ds}\zeta(s|A_f^2)|_{s=0}
\mp{i\pi\over 2}\eta(s|A_f)|_{s=0} \pm{i\pi\over
2}\zeta(s|A_f^2)|_{s=0}
\nonumber\\
& & = \biggl[-{1\over2}{d\over ds}\pm{i\pi\over2}\biggr]\biggl(
{1\over\Gamma(s)}\int_0^\infty dt
t^{s-1}Tr(e^{-tA_f^2})\biggr)|_{s=0}
\mp{i\pi\over2}{1\over\Gamma({s+1\over2})}\int_0^\infty dt
t^{{s+1\over2}-1}
Tr(A_f e^{-tA_f^2})|_{s=0},\nonumber\\
& &
\end{eqnarray}
where the $\mp$ sign in front of $\eta(0)$, a non-local object,
represents an ambiguity in the definition of the determinant. The
$\zeta(0|A_f^2)$ term can be reabsorbed into the contribution of
$\zeta^\prime(0|A_f^2)$, and hence will be dropped below. Here
$Tr\equiv\int dx\langle x|...|x\rangle\equiv\int dx tr(...)$. The
idea is that if one gets a match in the Seeley - de Witt
coefficients for the bosonic and fermionic determinants, implying
equality of UV-divergence, this is indicative of a possible complete
cancelation.

The heat kernel expansions for the bosonic and fermionic
determinants\cite{Gilkey}, in three dimensions, are given by:
\begin{eqnarray}
\label{eq:heatkernexps} & & tr(e^{-tA_b})=\sum_{n=0}^\infty
e_n(x,A_b) t^{{(n-3)\over2}}, tr(A_fe^{-tA_f^2})=\sum_{n=0}^\infty
a_n(x,A_f) t^{{(n-4)\over2}}.
\end{eqnarray}
For Laplace-type operators $A_b$ and $A_f^2$ (the $b$ implies
bosonic and $f$ implies fermionic), the non-zero coefficients in the
bulk, are determined to be the following:
\begin{eqnarray}
\label{eq:bosSdW}
\label{eq:e_ns} & & e_0(x,A_b)=(4\pi)^{-{3\over2}}Id,\
e_2(x,A_b)=(4\pi)^{-{3\over2}}\biggl[\alpha_1 E+\alpha_2R\
Id\biggr],
\end{eqnarray}
where $R$ is the Ricci scalar constructed from suitable pull-backs
of the metric and affine connection, $\alpha_i$'s are constants,
$Id$ is the identity that figures with the scalar leading symbol in
the Laplace-type operator $A_b$ (See \cite{Gilkey}), and
\begin{eqnarray}
\label{eq:Edieuf} & & E\equiv B -
G^{ij}(\partial_i\omega_j+\omega_i\omega_j-\omega_k\Gamma^k_{ij}),
\nonumber\\
& & A_b\equiv-(G^{ij}Id\partial_i\partial_j+A^i\partial_i+B),\nonumber\\
& & \omega_i={G_{ij}(A^j+G^{kl}\Gamma^j_{kl}Id)\over 2}.
\end{eqnarray}
For matrix-valued $E$, as will be the case for the Laplace-type
${\cal O}_1$ and the Dirac-type ${\cal O}_3$ in this paper, it is
understood that one has to take a trace.

From the expressions of ${\cal O}_{1,2,3}$, one sees that the
effective pullback of the metric (which gets used in, e.g.,
(\ref{eq:bosSdW}) and (\ref{eq:Edieuf})) on to the world volume of
the supersymmetric 3-cycle is given by:
$$G_{ij}=\frac{g_{ij}}{\sqrt{g}}=\left(\matrix{ G_{11} & 0 & G_{13}\cr
0 & G_{22} & 0 \cr
G_{13} & 0 & G_{23}\cr
}\right),$$ where the components are either ${\cal O}(1) +
{\cal O}(\zeta^2)$, e.g.,
$$G_{11}=\frac{{\sqrt{\left( 4 + e^{\frac{x}{2}} \right)
\,\left( 4 + e^x \right) }}}{2\,\left( 4 + e^x \right) } +
\frac{4\,\left( \frac{2\,{\tanh (z)}^2}{{\left( -4 + {{\rm sech}(z)}^2 \right) }^2} -
\frac{\frac{\left( 4 + e^{\frac{x}{2}} \right) \,{\tanh (z)}^2}{{\left( -4 + {{\rm sech}(z)}^2 \right) }^2} +
\left( 4 + e^x \right) \,\left( \frac{\left( 4 + e^{\frac{x}{2}} \right) \,{\left( 2 + \cosh (2\,z) \right) }^2}
{{\left( 1 + 2\,\cosh (2\,z) \right) }^4} + \frac{{\tanh (z)}^2}{{\left( -4 + {{\rm sech}(z)}^2 \right) }^2}
\right) }{4 + e^x} \right) \,{{\zeta}}^2}{e^x\,
{\sqrt{\left( 4 + e^{\frac{x}{2}} \right) \,\left( 4 + e^x \right) }}} +
{{O}({\zeta})}^3,$$ or ${\cal O}(\zeta^2)$, e.g.,

$$G_{13}=\frac{-8\,
\left( 2 + \cosh (2\,z) \right) \,\sinh (2\,z)\,{{\zeta}}^2}{e^x\,
{\sqrt{\left( 4 + e^{\frac{x}{2}} \right) \,\left( 4 + e^x \right) }}\,{\left( 1 + 2\,\cosh (2\,z) \right) }^3} +
{{O}({\zeta})}^3,$$

The corresponding vielbeins (which get used in the evaluation of
spin connections relevant to (\ref{eq:A1b})-(\ref{eq:A3b})) are
therefore given by:
$$e^{\ a^\prime}_i=\left(
\matrix{ 0 & e^{\ 2}_1 & e^{\ 3}_1\cr 0
&e^{\ 1}_2 & e^{\ 3}_2 \cr e^{\ 1}_3& 0 & 0 \cr }\right),$$
where all components are of ${\cal O}(1)+{\cal O}(\zeta^2)$, e.g.,

$$e^{\ 2}_1=\frac{e^{\frac{x}{2}}\,\cos (y)}
{{\sqrt{2}}\,{\left( \left( 4 + e^{\frac{x}{2}} \right) \,\left( 4 + e^x \right) \right) }^{\frac{1}{4}}} -
\frac{2\,{\sqrt{2}}\,\cos (y)\,\left( \frac{\left( 4 + e^{\frac{x}{2}} \right) \,{\tanh (z)}^2}
{{\left( -4 + {{\rm sech}(z)}^2 \right) }^2} +
\left( 4 + e^x \right) \,\left( \frac{\left( 4 + e^{\frac{x}{2}} \right) \,{\left( 2 + \cosh (2\,z) \right) }^2}
{{\left( 1 + 2\,\cosh (2\,z) \right) }^4} + \frac{{\tanh (z)}^2}{{\left( -4 + {{\rm sech}(z)}^2 \right) }^2}
\right) \right) \,{{\zeta}}^2}{e^{\frac{x}{2}}\,
{\left( \left( 4 + e^{\frac{x}{2}} \right) \,\left( 4 + e^x \right) \right) }^{\frac{5}{4}}}\hfill$$
$$ +
{{O}({\zeta})}^3,$$ and
$${\cal E}^{\ a^{\prime\prime}}_i=
\left(\matrix{ {\cal E}^{\ 1}_1 & {\cal E}^{\ 2}_1 & {\cal E}^{\
3}_1 & 0 \cr 0 & {\cal E}^{\ 2}_2 & {\cal E}^{\ 3}_2& {\cal E}^{\
4}_2 \cr 0 & {\cal E}^{\ 2}_3 & {\cal E}^{\ 3}_3 & 0 \cr }\right)$$
where the components are either of ${\cal O}(1)+{\cal O}(\zeta^2)$,
e.g.,
$${\cal E}^{\ 1}_1=-{\left( \left( 1 + \frac{e^{\frac{x}{2}}}{4} \right)
\,\left( 4 + e^x \right) \right) }^{-\left( \frac{1}{4} \right) } +
\frac{4\,{\sqrt{2}}\,\left( \frac{\left( 4 + e^{\frac{x}{2}} \right) \,{\tanh (z)}^2}
{{\left( -4 + {{\rm sech}(z)}^2 \right) }^2} +
\left( 4 + e^x \right) \,\left( \frac{\left( 4 + e^{\frac{x}{2}} \right) \,{\left( 2 + \cosh (2\,z) \right) }^2}
{{\left( 1 + 2\,\cosh (2\,z) \right) }^4} + \frac{{\tanh (z)}^2}{{\left( -4 + {{\rm sech}(z)}^2 \right) }^2}
\right) \right) \,{{\zeta}}^2}{e^x\,
{\left( \left( 4 + e^{\frac{x}{2}} \right) \,\left( 4 + e^x \right) \right) }^{\frac{5}{4}}} +
{{O}({\zeta})}^3,$$ or of ${\cal O}(\zeta)$, e.g.,

$${\cal E}^{\ 2}_1=\frac{-2\,\cos (y)\,\tanh (z)\,{\zeta}}
{e^{\frac{x}{2}}\,{\left( \left( 1 + \frac{e^{\frac{x}{2}}}{4} \right) \,\left( 4 + e^x \right) \right) }^{\frac{1}{4}}\,
\left( -4 + {{\rm sech}(z)}^2 \right) } + {{O}({\zeta})}^3.$$

The affine connection (relevant for evaluation of $\omega_i^b$ - see
(\ref{eq:w1b}) - which gets used in the evaluation of $e_n$s via
(\ref{eq:Edieuf}) and (\ref{eq:bosSdW})) for $G_{ij}$ are given as
under:
$$\Gamma^i_{jk}=\left(\matrix{
\Gamma^1_{11}&{{O}({\zeta})}^4&\Gamma^1_{13}
\cr
{{O}({\zeta})}^4&\Gamma^1_{22}&{{O}({\zeta})}^4\cr
{{O}({\zeta})}^4&\Gamma^1_{32}&\Gamma^1_{33} \cr
& & \cr
{{O}({\zeta})}^4&\Gamma^2_{12}&{{O}({\zeta})}^4\cr \Gamma^2_{21}&{{O}({\zeta})}^4&
\Gamma^2_{23}\cr
{{O}({\zeta})}^4&\Gamma^2_{32}
&{{O}({\zeta})}^4 \cr
& \cr
\Gamma^3_{11}&{{O}({\zeta})}^4&
\Gamma^3_{13}\cr
{{O}({\zeta})}^4&
\Gamma^3_{22}&
{{O}({\zeta})}^4\cr
\Gamma^3_{31}&{{O}({\zeta})}^4&
\Gamma^3_{33} \cr }\right).$$ Using
$\Gamma^i_{jk}$, one can then evaluate the various components of
the curvature tensor, the non-zero being: $R^2_{121}, R^3_{131}, R^2_{123},
R^3_{232},$ $ R^2_{323},R^4_{343}$. Using these, one evaluates the
non-zero components of the Ricci tensor: $R_{11}, R_{13}, R_{22}$
and $R_{33}$ which gives the Ricci scalar (relevant for evaluation
of the Seeley de-Witt coefficient $e_2(x,{\cal O}_{1\ \rm or\ 2}\
{\rm or}\ {\cal O}_3^2)$ - see (\ref{eq:bosSdW}) ):
$$R={\cal O}(1)+{\cal O}(\zeta^2)+{\cal O}(\zeta^4),$$ where the
first two terms, which can be easily calculated, are relevant to the
order we are working. As the actual expression for $R$ does not
explicitly get used in this paper and is also very long, we skip
giving the same.

The three-fold is topologically $M_3={\bf R}\times[0,1]\times S^1$,
implying that it has a boundary which is given by $\partial
M_3=(R\times S^1,0)\cup(R\times S^1,1)$. Note that configurations
involving branes wrapping noncompact cycles have been studied
earlier - see \cite{Kachruetal}. The three-fold is given by the set
of following equations:
\begin{eqnarray}
\label{eq:eqnssurf}
& & (x^4)^2+(x^5)^2=e^{-x^6},\nonumber\\
& & \frac{x^5}{x^4}=-tan(\frac{x^{10}}{2}),\nonumber\\
& & \frac{x^8}{x^7}=tan(\frac{x^{10}}{2}),\nonumber\\
& &
(x^7)^2+(x^8)^2=\frac{\zeta^2e^{x^6}(2-\frac{3x^8e^{-\frac{x^6}{2}}sec(\frac{x^{10}}{2})}{4\zeta}
-2\sqrt{1-\frac{3x^8e^{-\frac{x^6}{2}}sec(\frac{x^{10}}{2})}{4\zeta}})}{4\biggl(3-\frac{\biggl(2-\frac{3x^8e^{-\frac{x^6}{2}}sec(\frac{x^{10}}{2})}{4\zeta}
-2\sqrt{1-\frac{3x^8e^{-\frac{x^6}{2}}sec(\frac{x^{10}}{2})}{4\zeta}}\biggr)}{4}\biggr)^2}.
\end{eqnarray}
One notices the following ${\bf Z}_2$ symmetry of
(\ref{eq:eqnssurf}): \begin{equation} \label{eq:Z_2symmetry}
x_{4,7,6}\rightarrow x_{4,7,6};\ x_{5,8,10}\rightarrow-x_{5,8,10};\
\zeta\rightarrow-\zeta.
\end{equation}
Notice that under the above antiholomorphic involution: $J=du\wedge
d{\bar u} + dv\wedge d{\bar v} + ds\wedge d{\bar s}$ is reflected,
and $\Omega=du\wedge dv\wedge ds$ is complex conjugated. This is
related to the involution used in the construction of a $G_2$
manifold from a Calabi-Yau three-fold using the Joyce's
prescription: $\frac{CY_3\times S^1}{{\bf Z}_2}$.

Given that the supersymmetric three-fold $M_3$ has a nontrivial boundary, in addition to the bulk Seeley de-Witt
coefficients given by (\ref{eq:e_ns}), one also needs to evaluate boundary Seeley-de Witt coefficients. The latter for
$M_3$ of (\ref{eq:eqnssurf}) are given by (See \cite{Gilkey}):
\begin{eqnarray}
\label{eq:SdWboundaryM3} & & a_1^{\partial
M_3}=\frac{1}{4\pi}(\frac{\Gamma(\frac{3}{2})}{\Gamma(\frac{1}{2})\Gamma(2)}-1)tr{\bf
1},\nonumber\\
& & a_2^{\partial
M_3}=\frac{1}{(4\pi)^{\frac{3}{2}}}\frac{1}{3}(1-\frac{3}{4}\pi\frac{\Gamma(\frac{3}{2})}{\Gamma(\frac{1}{2})\Gamma(2)})
\Pi_{ii}|_{\partial M_3},
\end{eqnarray}
$\Pi_{ii}$ being the trace of the second fundamental form. The
second fundamental form is given by: $\Pi_{ij}=\langle
n,\bigtriangledown_{u_i}v_j\rangle$, where the tangent vectors
$u_i=\frac{\partial{x^m}}{\partial y^i}\partial_m$, and similarly
for $v_j$, where $m$ takes values $3,6,10$. For $\partial M_3$,
$M_3$ being given by (\ref{eq:eqnssurf}),
$\Pi_{ij}=n^{x^6}\frac{\partial^2 x^6}{\partial y_i\partial y_j}$,
which for the embedding $x^6=-y^1$, vanishes.

We now proceed with the evaluation of the bulk Seeley de-Witt
coefficients as given in (\ref{eq:e_ns}). For this we would first
evaluate $A^i_b$s (the ``$b$" implies relevant to bosonic operators)
utilizing the results for the vielbeins obtained earlier in this
section. This is done in equations (\ref{eq:A1b})-(\ref{eq:A3b}).
Using the results for $A^i_b$s, we would then calculate
$\omega_i^b$s. The same is done in equation (\ref{eq:w1b}) for
$\omega_1^b$ - one can similarly evaluate $\omega_2^b$ and
$\omega_3^b$. We would then be able to calculate $E$, which would
enable us to evaluate $e_n$s. This is done in equations
(\ref{eq:E1b})-(\ref{eq:trEb}).

For the operator ${\cal O}_1$, the expressions for $A^i_b$ (see
(\ref{eq:bosondets}) and (\ref{eq:Edieuf}) ) are given as:
\begin{eqnarray}
\label{eq:A1b} & &
A^1_b=\delta_{uv}G^{1j}(\omega_j|_{M_3} +
\omega_j|_{N(M_3)\hookrightarrow X_{G_2}})\nonumber\\
& & =\left(\matrix{ 0 & a_{12} & a_{13}& 0 & 0 & 0 & 0 \cr -a_{12}
& 0 & a_{23} & 0 & 0 & 0 & 0 \cr -a_{13} & -a_{23} & 0 & 0 & 0 & 0 &
0 \cr
0 & 0 & 0 & 0 & a_{45}
& a_{46} & {{O}(
\zeta)}^2 \cr
0 & 0 & 0 & -a_{45} & 0 & a_{56}& a_{57} \cr
0 & 0 & 0 & -a_{46} & -a_{56} & 0 & a_{67} \cr 0 & 0 & 0 & {{O}(
\zeta)}^2 & -a_{57} & -a_{67} & 0 \cr }\right)\end{eqnarray}
($u,v$ index ${\bf R}^4$-valued coordinates and $N(M_3)$ is the
normal bundle to $M_3$). Similarly,
\begin{eqnarray}
\label{eq:A2b} & & A^2_b=\delta_{uv}G^{2j}(\omega_j|_{M_3} +
\omega_j|_{N(M_3)\hookrightarrow X_{G_2}})\nonumber\\
& & =\left(\matrix{ 0 & b_{12} & b_{13}& 0 & 0 & 0 & 0\cr
-b_{12} & 0 & b_{23} & 0 & 0 & 0 & 0 \cr
-b_{13} & -b_{23}& 0 & 0 & 0 & 0 & 0 \cr
0 & 0 & 0 & 0 & b_{45}& b_{46}& b_{47}\cr
0 & 0 & 0 & -b_{45} & 0 & b_{56}& b_{57}\cr
0 & 0 & 0 & -b_{46} & -b_{56} & 0 & b_{67} \cr
0 & 0 & 0 & -b_{47} & -b_{57} & -b_{67} & 0 \cr }\right),
\end{eqnarray}
and
\begin{eqnarray}
\label{eq:A3b} & &
A^3_b=\delta_{uv}G^{3j}(\omega_j|_{M3}+\omega_j|_{N(M_3)\hookrightarrow
X_{G_2}})\nonumber\\
& & =\left(\matrix{ 0 & c_{12} & c_{13}
& 0 &
0 & 0 & 0 \cr
-c_{12} & 0 &
c_{23} & 0 & 0 & 0 & 0 \cr
-c_{13}& -c_{23} & 0 & 0 & 0 & 0 & 0 \cr
0 & 0 & 0 & 0 & c_{45}& c_{46} & {{O}(\zeta)}^2 \cr
0 & 0 & 0 & -c_{45}& 0 & c_{56}
&
{{O}(\zeta)}^2 \cr
0 & 0 & 0 & -c_{46} & -c_{56} & 0 & {{O}(\zeta)}^2 \cr
0 & 0 & 0 & {{O}(\zeta)}^
2 & {{O}(\zeta)}^2 & {{O}(\zeta)}^2 & 0 \cr }\right),
\end{eqnarray}
where the various non-zero elements can easily be worked out.

To evaluate the $\zeta$ Seeley de-Witt coefficients, one needs to
evaluate $\omega^b_i$. The expressions for $\omega_i^b$ are given
as: \begin{eqnarray} \label{eq:w1b} & &
\omega_1^b=\frac{G_{1j}}{2}(A^j_b + G^{kl}\Gamma^j_{kl}{\bf
1}_7)\nonumber\\
& &=\left(\matrix{ w_{111} & w_{112}& w_{113} & {{O}(\zeta)}^4 & {
{O}(\zeta)}^4 & {{O}(\zeta)}^4 & {{O}(\zeta)}^4 \cr
w_{121} & w_{122} & w_{123} & {{O}(\zeta)}^4 & {{O}(\zeta)}^4 & {{O}(
\zeta)}^4 & {{O}(\zeta)}^4 \cr
w_{131} & w_{132} & w_{133} & {{O}(\zeta)}^4 & {
{O}(\zeta)}^4 & {{O}(\zeta)}^4 & {{O}(\zeta)}^4 \cr
{{O}(\zeta)}^4 & {{O}(\zeta)}^4 & {{O}(\zeta)}^4 & w_{144}
& w_{145}& w_{146}& {{O}(\zeta)}^2 \cr {{O}(\zeta)}^4 & {{O
}(\zeta)}^4 & {{O}(\zeta)}^4 & w_{154} & w_{155} & w_{156} & w_{157} \cr
{{O}(\zeta)}^4 & {{O}(\zeta)}^4 & {{O}(\zeta)}^4 & w_{164}
& w_{165} & w_{166}& w_{167}\cr
{{O}(\zeta)}^4 & {{O}(\zeta)}^4 & {{O
}(\zeta)}^4 & {{O}(\zeta)}^2 & w_{175} & w_{176}& w_{177}
\cr })\right)\end{eqnarray}
One can similarly evaluate expressions for $\omega_{2,3}^b$, using
which one can evaluate $E({\cal O}_1)=E_1^b+E_2^b+E_3^b$, where
\begin{eqnarray}
\label{eq:E1b} & & E_1^b=-G^{ij}\partial_i\omega_j^b\nonumber\\
& &
=\left(\matrix{E_{111}
& E_{112}
&E_{113}&{{O}(\zeta)}^4&{{O}(\zeta)}^4&
{{O}(\zeta)}^4&{{O}(\zeta)}^4\cr
E_{121}&E_{122}&E_{123}
&{{O}(\zeta)}^4&{{O}(\zeta)}^4&
{{O}(\zeta)}^4&{{O}(\zeta)}^4\cr
E_{131}& E_{132}
&E_{133}&{{O}(\zeta)}^4&
{{O}(\zeta)}^4&{{O}(\zeta)}^4&{{O}(\zeta)}^4\cr
{{O}(\zeta)}^4&{{O}(\zeta)}^4&{{O}(\zeta)}^4&
E_{144} &
E_{145}& E_{146}
&{{O}(\zeta)}^2\cr
{{O}(\zeta)}^4&
{{O}(\zeta)}^4&{{O}(\zeta)}^4&
E_{154} & E_{155}
&
E_{156} & E_{157}\cr
{{O}(\zeta)}^4&{{O}(\zeta)}^4&{{O}(\zeta)}^4& E_{164}
& E_{165}
&E_{166}& E_{167}\cr
{{O}(\zeta)}^4&{{O}(\zeta)}^4&{{O}(\zeta)}^4&
{{O}(\zeta)}^2&E_{175}&E_{176}
& E_{177} \cr}\right),
\end{eqnarray}
where the non-zero elements can be easily worked out. Similarly,
\begin{eqnarray}
\label{eq:E2b} & & E_2^b=-G^{ij}\omega^b_i\omega_j^b\nonumber\\
& &
=\left(\matrix{ E_{211}& E_{212}
& E_{213}
&
{{O}(\zeta)}^8 & {{O}(\zeta)}^8 & {{O}(\zeta)}^8 &
{{O}(\zeta)}^8\cr
E_{221} & E_{222}
& E_{223}
& {{O}(\zeta)}^8 & {{O}(\zeta)}^8 &
{{O}(\zeta)}^8 & {{O}(\zeta)}^8\cr
E_{231} & E_{232}
&E_{233} &
{{O}(\zeta)}^8 & {{O}(\zeta)}^8 & {{O}(\zeta)}^8 &
{{O}(\zeta)}^8\cr
{{O}(\zeta)}^8 & {{O}(\zeta)}^8 &
{{O}(\zeta)}^8 & E_{244} & E_{245}
& E_{246}& E_{247} \cr
{{O}(\zeta)}^8 & {{O}(\zeta)}^8 &
{{O}(\zeta)}^8 & E_{254} & E_{255} &
{{O}(\zeta)}^4 & E_{257}\cr
{{O}(
\zeta)}^8 & {{O}(\zeta)}^8 & {{O}(\zeta)}^8 &
E_{264} & {{O}(\zeta)}^4 &
E_{266} &
E_{267}\cr
{{O}(
\zeta)}^8 & {{O}(\zeta)}^8 & {{O}(\zeta)}^8 &
E_{274} & E_{275} &
E_{276} & E_{277}
\cr}\right)
\end{eqnarray}
and \begin{eqnarray} \label{eq:E3b} & &
E_3^b=G^{ij}\omega_k^b\Gamma^k_{ij}\nonumber\\
& & =\left(\matrix{ E_{311} & E_{312} & E_{313} & {{O}(\zeta)}^4 &
{{O}(\zeta)}^4 & {{O}(\zeta)}^4 & {{O}(\zeta)}^4\cr
E_{321} &
E_{322} &E_{323} & {{O}(\zeta)}^4 & {{O}(\zeta)}^4 &
{{O}(\zeta)}^4 & {{O}(\zeta)}^4\cr
E_{331} & E_{332}
& E_{333} & {{O}(\zeta)}^4 & {{O}(\zeta)}^4 &
{{O}(\zeta)}^4 & {{O}(\zeta)}^4\cr
{{O}(\zeta)}^4 &
{{O}(\zeta)}^4 & {{O}(\zeta)}^4 & E_{344}
& E_{345} &E_{346} &
{{O}(\zeta)}^2\cr
{{O}(\zeta)}^4 & {{O}(\zeta)}^4 &
{{O}(\zeta)}^4 & E_{354} &
E_{355} & E_{356} & E_{357} \cr
{{O}(\zeta)}^4 & {{O}(\zeta)}^4 & {{O}(\zeta)}^4 &
E_{364} & E_{365} & E_{366}
& E_{367}\cr
{{O}(\zeta)}^4 & {{O}(\zeta)}^4 & {{O}(\zeta)}^4 &
{{O}(\zeta)}^2 & E_{375}&
E_{376} & E_{377}\cr}\right).
\end{eqnarray}

One thus gets:
\begin{eqnarray}\label{eq:trEb}
& & {\rm tr}(E({\cal O}_1))={\rm tr}(E_1^b+E_2^b+E_3^b)= \frac{( 4 +
e^x ) \,( -1536\,e^{\frac{x}{2}} + 2192\,e^x +
64\,e^{\frac{3\,x}{2}} + 216\,e^{2\,x} - 112\,e^{\frac{5\,x}{2}} +
e^{3\,x} ) }{128\,{( ( 4 + e^{\frac{x}{2}} ) \,
( 4 + e^x ) ) }^{\frac{5}{2}}} + \nonumber\\
& &
\frac{E \zeta^2}{128\,e^x\,
{( ( 4 + e^{\frac{x}{2}} ) \,( 4 + e^x )
) }^{\frac{7}{2}}\,{( 1 + 2\,\cosh (2\,z) ) }^6} +
{O}({\zeta}^3),
\end{eqnarray}
where
\begin{eqnarray}
\label{eq:Edieuf1} & & E\equiv ( 4 + e^x ) \, \biggl( 4237426688 +
4268851200\,e^{\frac{x}{2}} + 4797884672\,e^x +
3473765744\,e^{\frac{3\,x}{2}}\nonumber\\
& & + 2023406480\,e^{2\,x} +
1003298120\,e^{\frac{5\,x}{2}} + 381310232\,e^{3\,x} +
117349175\,e^{\frac{7\,x}{2}} + 28347453\,e^{4\,x}\nonumber\\
& & +
4231472\,e^{\frac{9\,x}{2}} + 267264\,e^{5\,x} +
4\,( 617218048 + 628211712\,e^{\frac{x}{2}} + 705386752\,e^x\nonumber\\
& & +
512375632\,e^{\frac{3\,x}{2}} + 299013040\,e^{2\,x} +
148217272\,e^{\frac{5\,x}{2}} + 56510632\,e^{3\,x} +
17391397\,e^{\frac{7\,x}{2}} + 4212567\,e^{4\,x}\nonumber\\
& & +
635494\,e^{\frac{9\,x}{2}} + 40704\,e^{5\,x} ) \,\cosh (2\,z) -
4\,( 554958848 + 551976960\,e^{\frac{x}{2}} + 620804864\,e^x +
447750848\,e^{\frac{3\,x}{2}} + 260136512\,e^{2\,x}\nonumber\\
& & +
129088208\,e^{\frac{5\,x}{2}} + 48864080\,e^{3\,x} +
15043796\,e^{\frac{7\,x}{2}} + 3621852\,e^{4\,x} +
532883\,e^{\frac{9\,x}{2}} + 33024\,e^{5\,x} ) \,\cosh (4\,z)\nonumber\\
& & -
726663168\,\cosh (6\,z) - 733347840\,e^{\frac{x}{2}}\,\cosh (6\,z) -
821714944\,e^x\,\cosh (6\,z) -
596084096\,e^{\frac{3\,x}{2}}\,\cosh (6\,z)\nonumber\\
& & -
346978944\,e^{2\,x}\,\cosh (6\,z) -
172383936\,e^{\frac{5\,x}{2}}\,\cosh (6\,z) -
65568064\,e^{3\,x}\,\cosh (6\,z)\nonumber\\
& & -
20221080\,e^{\frac{7\,x}{2}}\,\cosh (6\,z) -
4890248\,e^{4\,x}\,\cosh (6\,z) -
728600\,e^{\frac{9\,x}{2}}\,\cosh (6\,z)\nonumber\\
& & -
46080\,e^{5\,x}\,\cosh (6\,z) - 25165824\,\cosh (8\,z) -
30932992\,e^{\frac{x}{2}}\,\cosh (8\,z) - 32834560\,e^x\,\cosh (8\,z) -
25648768\,e^{\frac{3\,x}{2}}\,\cosh (8\,z)\nonumber\\
& & -
15232384\,e^{2\,x}\,\cosh (8\,z) -
7692928\,e^{\frac{5\,x}{2}}\,\cosh (8\,z) -
3094272\,e^{3\,x}\,\cosh (8\,z) -
975624\,e^{\frac{7\,x}{2}}\,\cosh (8\,z)\nonumber\\
& & -
247896\,e^{4\,x}\,\cosh (8\,z) -
41180\,e^{\frac{9\,x}{2}}\,\cosh (8\,z) - 3072\,e^{5\,x}\,\cosh (8\,z) +
2621440\,\cosh (10\,z) + 2031616\,e^{\frac{x}{2}}\,\cosh (10\,z)\nonumber\\
& & +
2505728\,e^x\,\cosh (10\,z) +
1632320\,e^{\frac{3\,x}{2}}\,\cosh (10\,z) +
924608\,e^{2\,x}\,\cosh (10\,z) +
455648\,e^{\frac{5\,x}{2}}\,\cosh (10\,z)\nonumber\\
& & +
149152\,e^{3\,x}\,\cosh (10\,z) +
43780\,e^{\frac{7\,x}{2}}\,\cosh (10\,z) +
8748\,e^{4\,x}\,\cosh (10\,z) + 960\,e^{\frac{9\,x}{2}}\,\cosh (10\,z)\nonumber\\
& & -
131072\,\cosh (12\,z) - 98304\,e^{\frac{x}{2}}\,\cosh (12\,z) -
110848\,e^x\,\cosh (12\,z) - 63984\,e^{\frac{3\,x}{2}}\,\cosh (12\,z) -
32528\,e^{2\,x}\,\cosh (12\,z)\nonumber\\
& & -
12296\,e^{\frac{5\,x}{2}}\,\cosh (12\,z) -
5464\,e^{3\,x}\,\cosh (12\,z) -
1535\,e^{\frac{7\,x}{2}}\,\cosh (12\,z) - 469\,e^{4\,x}\,\cosh (12\,z)
\biggr).
\end{eqnarray}
Despite the very complicated and long (\ref{eq:trEb}) and
(\ref{eq:Edieuf1}), we will see shortly that one gets a remarkable
result, which is that the square of the relevant Dirac-type
operator, ``$A_f^2$" (``$f$" denoting fermionic) - ${\cal O}_3^2$ -
contributes {\it precisely\ } as $``A_b"(\equiv{\cal O}_1)$ - see
(\ref{eq:check}). This, in itself is a check of our lengthy spectral
analysis, because (\ref{eq:check}) and the fact (which again we show
momentarily) that the $\eta$-function contribution from $A_f$
vanishes, is something we had anticipated from supersymmetry
arguments (given that one is dealing with a supersymmetric
three-fold for getting the membrane instanton), but the same was
totally unobvious from a spectral analysis point of view.

We now do a heat-kernel asymptotics analysis of the fermionic
determinant $det{\cal O}_3$. The fermionic operator ${\cal O}_3$ can
be expressed as:
\begin{eqnarray}
\label{eq:O3} & & {\cal O}_3\equiv
\sqrt{g}g^{ij}\Gamma_jD_i=\sqrt{g}g^{ij}\Gamma_j
\biggl(\partial_i+{1\over4}\omega^{a^\prime b^\prime}_i
\Gamma_{a^\prime b^\prime} +{1\over4}\omega^{a^\prime
b^{\prime\prime}}_i\Gamma_{a^\prime b^{\prime\prime}} \biggr) \equiv
G^{ij}\Gamma_j\partial_i-r,
\end{eqnarray}
where
\begin{eqnarray}
\label{eq:Grdieufs} & & G^{ij}\equiv\sqrt{g}g^{ij};
r\equiv{-1\over4}\sqrt{g}g^{ij}\Gamma_j\biggl( \omega^{a^\prime
b^\prime}_i\Gamma_{a^\prime b^\prime} +\omega^{a^\prime
b^{\prime\prime}}_i\Gamma_{a^\prime b^{\prime\prime}}\biggr),
\end{eqnarray}
and using the results of section {\bf 3}, we set $\omega^{a^\prime
b^{\prime\prime}}_i=0$. ${\cal O}_3$ is of the Dirac-type as ${\cal
O}_3^2$ is of the Laplace-type, as can be seen from the following:
\begin{eqnarray}
\label{eq:O3squared} & & {\cal O}_3^2\equiv
G^{ij}\partial_i\partial_j+A^i\partial_i+B,\ {\rm where}:
\nonumber\\
& & G^{ij}\equiv \sqrt{g}g^{ij};\nonumber\\
& & A^i\equiv G^{lk}\Gamma_l\partial_k(G^{ji}\Gamma_j) + G^{ji}G^{lk}\Gamma_j\Gamma_l\omega_k+G^{kl}G^{ji}\Gamma_l\omega_k\Gamma_j;\nonumber\\
& & B\equiv
G^{ij}\Gamma_i\partial_j(G^{kl}\Gamma_k\omega_l)+G^{ij}G^{kl}\Gamma_i\omega_j\Gamma_k\omega_l.
\end{eqnarray}
The remark regarding the {\it dissimilar} ${\cal O}_1$ and ${\cal
O}_3^2$ in the introduction is justified by comparing
(\ref{eq:bosondets}) and (\ref{eq:O3squared}). Now,
\begin{equation}
\label{eq:phi1} {\cal O}_3\equiv G^{ij}\Gamma_j\bigtriangledown_i
-\phi,
\end{equation}
where $\phi\equiv r+\Gamma^i\omega_i$, and
\begin{eqnarray}
\label{eq:omegacondieuf} & &
\omega_l\equiv{G_{il}\over2}(-\Gamma^j\partial_j\Gamma^i+\{r,\Gamma^i\}+G^{jk}\Gamma^i_{jk}).
\end{eqnarray}

The bulk Seeley-de Witt coefficients $a_i$ are given by (See
\cite{Gilkey}):
\begin{eqnarray}
\label{eq:fermionais} & & a_1(x,G^{ij}\Gamma_j\bigtriangledown_i
-\phi)=-(4\pi)^{-{3\over2}}tr(\phi);\nonumber\\
& & a_3(x, G^{ij}\Gamma_j\bigtriangledown_i
-\phi)=-{1\over6}(4\pi)^{-{3\over2}} tr(\phi R +6\phi{\cal
E}-\Omega_{a^\prime b^\prime;a^\prime} \Gamma_{b^\prime}),
\end{eqnarray}
where
\begin{equation}
\label{eq:cal Edieuf2} {\cal
E}\equiv-{1\over2}\Gamma^i\Gamma^j\Omega_{ij}+\Gamma^i\phi_{;i}-\phi^2,\
\Omega_{ij}\equiv\partial_i\omega_j-\partial_j\omega_i+[\omega_i,\omega_j],
\end{equation}
and $\Omega_{a^\prime b^\prime}=e_{a^\prime}^{\ i} e_{b^\prime}^{\
j}\Omega_{ij}$. To ensure that ${\cal O}_3^2$ is a Laplace-type
operator,
$\Gamma_i\equiv\frac{1}{g^{\frac{1}{4}}}\partial_iX^M\Gamma_M$, $M$
indexing the eleven (Euclidean) dimensions and $\Gamma_M$ being the
generators of $Cl(0,11)$. The boundary $\eta$-function Seeley-de
Witt coefficients (See \cite{Gilkey}) are given as:
\begin{eqnarray} & & a_2^{\partial
M_3}=\frac{1}{4\pi}(-\frac{1}{4}(\frac{\Gamma(\frac{3}{2})}{\Gamma(\frac{1}{2})\Gamma(2)}
- 1)\phi -
\frac{\frac{\Gamma(\frac{3}{2})}{\Gamma(\frac{1}{2})\Gamma(2)}}{4}\Gamma^i\omega_i)|_{\partial
M_3},\nonumber\\
& & a_3^{\partial
M_3}=\frac{-1}{3(4\pi)^{\frac{3}{2}}}\phi_{;3}|_{\partial M_3}.
\end{eqnarray}
Using the generators of $Cl(0,7)$:
\begin{eqnarray*}
& & \gamma_1=i\sigma^2\otimes\sigma^1\otimes{\bf 1}_2,\nonumber\\
& & \gamma_2=i\sigma^2\otimes \sigma^3\otimes{\bf 1}_2,\nonumber\\
& & \gamma_3={\bf 1}_2\otimes i\sigma^2\otimes\sigma^1,\nonumber\\
& & \gamma_4={\bf 1}_2\otimes i\sigma^2\otimes\sigma^3,\nonumber\\
& & \gamma_5=\sigma^1\otimes{\bf 1}_2\otimes i\sigma^2,\nonumber\\
& & \gamma_6=\sigma^3\otimes{\bf 1}_2\otimes i\sigma^2,\nonumber\\
& & \gamma_7=i\sigma^2\otimes i\sigma^2\otimes i\sigma^2,
\end{eqnarray*}
one can construct generators of $Cl(0,11)$ (See \cite{harvmoore})
as:
\begin{eqnarray}
\label{eq:Cl11} & & \Gamma_{a^\prime}=\sigma^{a^\prime}\otimes
(-\sigma^3)\otimes{\bf
1}_8,\nonumber\\
& & \Gamma_{a^{\prime\prime}}={\bf 1}_2\otimes
i\sigma^2\otimes\gamma_{a^{\prime\prime}},
\end{eqnarray}
One then sees that all the terms in the $\eta$ Seeley de-Witt
coefficients for ${\cal O}_3$ are of the type tr(odd,even) +
tr(even,odd) where one counts the (number of $\Gamma_{a^\prime}$'s,
number of $\Gamma_{a^{\prime\prime}}$'s). Now, using
(\ref{eq:Cl11}), one can show that
$tr(\prod_{i=1}^{2m+1}\Gamma_{a_i^{\prime}})=tr(\prod_{i=1}^{2n+1}\prod_{j=1}^{2m}\Gamma_{a^{\prime\prime}})=0$.
This implies that the bulk and boundary $\eta$ Seeley de-Witt
coefficients for ${\cal O}_3$ vanish.

Further, one sees that $B$ is traceless. Analogous to the bosonic
sector contribution, one can evaluate $A^i_f$ ($f$ denoting
fermionic contribution) and therefore calculate $\omega_i^f$'s, and
using the latter, one gets the incredible result
\begin{equation}
\label{eq:check} {\rm tr}(E({\cal O}_3))= {\rm tr}(E({\cal
O}_1^2))=(\ref{eq:trEb})\&(\ref{eq:Edieuf1})!!!
\end{equation}
From equations (\ref{eq:trEb}) and (\ref{eq:Edieuf1}), we see that
we get a match for the Seeley de-Witt coefficients, for terms {\it
including} ${\cal O}(\zeta^2)$ - in fact the non-triviality of the
calculations seem to be the perfect match of ${\cal O}(\zeta^2)$
terms for the bosonic and fermionic fluctuations. From
(\ref{eq:SU2ans1}), one sees that the dependence of the embedding of
the associative three-fold in the $G_2$-manifold is via the
dependence of the same on $\zeta$ - setting $\zeta$ to zero is
equivalent to the reduction of the world-volume integral $\int
d^3z(...)(s)=\int dx dy dz (...)(x,y,z)$ to the world-line integral
$\int dz(...)$ corresponding to the $D0$-brane of type $IIA$ theory
in the vanishing $M$-theory circle limit.

Further, using (\ref{eq:bosfermdets}), one thus conjectures that:
\begin{equation}
\frac{{\rm ln\ det}{\cal O}_3}{{\rm ln\ det}{\cal O}_1}=\frac{1}{2},
\end{equation}
implying that the noncompact instanton has a residual supersymmetry
- {\it arrived upon from a heat kernel asymptotics/spectral analysis
point of view}.

\section{Conclusion}

The Seeley de-Witt coefficients associated with the nonperturbative
superpotential generated by an MQCD-like instanton configuration
obtained by wrapping $M2$-brane around a noncompact supersymmetric
three-fold embedded in a (noncompact) $G_2$-manifold relevant to
MQCD, understood as the $M$ theory configuration dual to a type
$IIB$ configuration compactified on a circle of vanishing radius,
was considered in this paper\footnote{For a noncompact membrane instanton, what is
more appropriate to be considered is $e^{\frac{1}{l_{11}^3}vol(g)}\Delta W$
rather than $\Delta W$ - the former will be independent of the volume of the
noncompact instanton.}. The boundary $\eta$ Seeley de-Witt
coefficients for the relevant fermionic operator vanish. Up to
second order in a complex parameter that is part of the embedding of
the aforementioned three-fold in the $G_2$ seven-fold, we get a
perfect match between the Seeley de-Witt coefficients between the
fermionic and one of the two bosonic determinants thereby strongly
suggesting the presence of the {\it expected} surviving
supersymmetry of the nonperturbative configurations in $M$-theory.
From a spectral analysis point of view, the results themselves
provide a remarkable check - in particular, if one looks at the
extremely long and complicated expressions given in equations
(\ref{eq:trEb}) and (\ref{eq:Edieuf1}) for the Laplace-type operator
${\cal O}_1$, it is extremely nontrivial to see that one gets
exactly the same expression for the Dirac-type operator ${\cal O}_3$
in equation (\ref{eq:check}). Further, this also shows that one
might get quantum corrections from the uncancelled ln det ${\cal
O}_2$ (at least in the static gauge used). One has also to
appreciate that the quantities involved in the calculations, are not
just pullback of the space-time metric and the Gamma matrices, but
involve, e.g., pseudo-metrics (because of the extra square root of
the pulled back metric).

Given the
direct-product topology $S\times[0,1]$ of the $M2$-brane, one can ask the
question what happens if the $M2$-brane does in fact end on $M5$-branes
on the interval, or even $M9$-branes. One would then have to deal with the
contribution to the superpotential coming from the $M5-M5$, $M5-M9$ and
$M9-M9$ open membrane instantons - the $M9-M9$ instantons, the $M$-theory
analogues of world-sheet instantons, often sum up to zero (See \cite{Mooreetal}
and references therein) however. A sketch of the relevant expressions in the
context of heterotic $M$-theory is given in, e.g., \cite{Mooreetal}.
In the context of plain $M$-theory on $G_2$-manifolds, the $M5-M5$
superpotential in the supegravity approximation,
e.g., would be of the form: $e^{(X_1 - X_2)
\int_S(i_{\frac{\partial}{\partial X}}C+iJ)}(...)$, where $X_i$ is the
complexified position of the $M5$-branes obtained from the $M$-theory chiral
two form (corresponding to a self-dual field strength on the $M5$-brane
world-volume) - See \cite{harvmoore}. Based on arguments given in
\cite{Mooreetal}, one would guess (especially for ``barely'' $G_2$ manifolds)
that supersymmetry requirements would be met if the $M$-theory circle direction
is an appropriate function of the interval coordinate ($x$ in our paper),
and the other internal coordinates depend on $y$ and $z$ (of our paper).

The nonperturbative membrane instanton contribution to the
superpotential can be compared with the complexified
affine-Toda-like superpotential, generated by three-dimensional
instantons (or four-dimensional monopoles) in the compactification
of the $D=4, {\cal N}=1$ SYM on a circle to $D=3 ( {\cal N}=2$ SYM),
given by: $W\sim e^{-V} + e^{2i\pi\tau}e^{V}$ ($\tau\equiv\frac{4\pi
i}{g^2}+\frac{\theta}{2\pi}$), where the complex field $V$, formed
from the Wilson line for the gauge field along the circle and the
scalar dual to the three-dimensional gauge field, parametrize an
${\cal N}=2$ K\"{a}hler moduli space $\frac{T^2}{S^1}$ (See
\cite{affetalritzetal}).

The spirit of the paper is similar to the work of, e.g.,
Sonnenschein et al, in the late nineties - \cite{S} - on seeing
whether or not the classical Wilson loop in an $AdS_5\times S^5$
background, received quantum corrections. In these papers, the
authors provide examples of models where the authors explicitly
check whether or not one gets a cancelation between the bosonic and
the fermionic determinants implying whether or not the classical
result for the Wilson loop, receives quantum corrections.

To the best of our knowledge, a spectral/heat kernel asymptotics
analysis (based largely on the results in mathematics of Branson,
Gilkey and Kirsten) for membrane instantons obtained from a
supersymmetric three-fold with boundary, embedded in a
$G_2$-manifold, has never been worked out, and all the formulae used
in this paper are extremely useful not only in the context of
membrane instanton superpotential but also quantum corrections to
Wilson loops/surfaces.

\section*{Acknowledgements}

One of us(AM) would like to thank K.Ray, P.Ramadevi, A.Srivastava and specially A.P.Balachandran for
useful correspondences.

\end{document}